\documentclass{elsart}
\usepackage[final]{epsfig}
\usepackage{amsmath}
\usepackage{graphics}

\newcommand{\be}{\begin{equation}}
\newcommand{\ee}{\end{equation}}
\newcommand{\beas}{\begin{eqnarray*}}
\newcommand{\eeas}{\end{eqnarray*}}
\newcommand{\bea}{\begin{eqnarray}}
\newcommand{\eea}{\end{eqnarray}}

\begin{document} 

\begin{frontmatter}

\title{Exploring an opinion network for taste prediction: an empirical study}

\author {Marcel Blattner, Yi-Cheng Zhang}
\address{Department of Physics, University of Fribourg, Chemin du Musée 3, CH
  - 1700 Fribourg, Switzerland}
\author {Sergei Maslov}
\address{Department of Physics, Brookhaven National Laboratory, Upton, New
  York 11973}
\thanks[SNFS]{Supported by the Swiss National Science Foundation}

\begin{abstract}
We develop a simple statistical method to find affinity relations in
a large opinion network which is represented by a very sparse matrix. These relations allow us to predict
missing matrix elements. We test our method on the
\emph{Eachmovie} data of thousands of movies and viewers. We found
that significant prediction precision can be achieved and it is rather
stable. There is an intrinsic limit to further improve the prediction precision by collecting more
data, implying perfect prediction can never obtain via statistical means.
\end{abstract}


\end{frontmatter}

\section{Introduction}
With the advent of the World Wide Web (WWW) we witness the onset of what is often called 'Information Revolution'.
With so many sources and users linked together instantly we face both challenges and opportunities, specially for scientists.
The most prominent challenge is information overload: no one can possibly check out all the information potentially relevant for him. The most promising opportunity is that the WWW offers possibility to infer or deduce other users experience to indirectly boost a single user's information capability. Both computer scientists and internet entrepreneurs extensively use various collaborative-filtering tools to tap into this opportunity.   

The so-called web2.0 represents a new wave in web applications: many newer web
sites allow users' feedback, enable their clustering and communication. 
Much of users' feedback can be interpreted as votes or evaluation on the information sources. Such voting is much more widespread: our choice of movies, books,
consumer products and services could be considered as our votes
representing our tastes. With a view to develop a prediction-model suitable
for web application, we need to first test a model is a limited setting. 
For a more concrete example consider opinions of movie-viewers on
movies they have seen. We use in this work the \emph{EachMovie} dataset,
generously provided by the \emph{Compaq} company. The \emph{Eachmovie} dataset comprises
ratings on $1628$ movies by $72916$ users. The dataset has a density
of approximately $3\%$, meaning that $97\%$ of possible ratings are
absent. This dataset can represented by an information matrix: each user has only seen a
tiny fraction of all the movies; each movie has been seen by a large
number of users but they are only a tiny fraction of all users.
This (sparse) information matrix has $97\%$ elements missing; our task is to
find whether we can predict them leveraging affinity relations
hidden in the dataset.

\section{Prediction Algorithm and Results}
There is a particular way how such information on movies could be
used to recommend other users movies they have not yet seen but
which would likely suit their tastes. Such recommendations can be
made by a centralized agent (matchmaker) who collects a large
number of votes. The idea behind such services (called ``recommender
system'' or ``collaborative filtering'' by computer scientists
\cite{Res94}\cite{Breese98}\cite{Billsus98}\cite{Sarwar01} ) is that
users' votes are first used to measure the affinity of users' tastes.
Then opinions of users with tastes sufficiently
similar to the user in question are summed up to predict the opinion
on movies she/he has not seen yet. 
The data of the ``matchmaker'' are stored in the voting matrix $V$
with entries $v_{i\alpha}$, this is the vote of user $i$ to movie
$\alpha$. For simplicity we only take into account from the original
data users who have seen at least $200$ movies. As a further
approximation we shall compress the original votes $(1 to 5)$ to
$v_{i\alpha} \in \{-1,1\}$, i.e, $1$ and  $2$ are converted to
$-1$ (dislike), $4$ and $5$ to $1$ (like), $3$ is interpreted as 0, as if the
user has not seen the movie. Elsewhere we show that such simplifying
approximations do not induce statistically significant reduction in
prediction power. The dimension of the rectangular matrix $V$ is
$(1223 \times 1648)$, i.e. there are $N=1223$ users and $M=1648$
movies. In this matrix there are $\sum_{i,\alpha}|v_{i\alpha}| \sim
2 \cdot 10^{5}$ non-zero elements (votes).

\begin{figure}
\includegraphics[width=0.9 \textwidth]{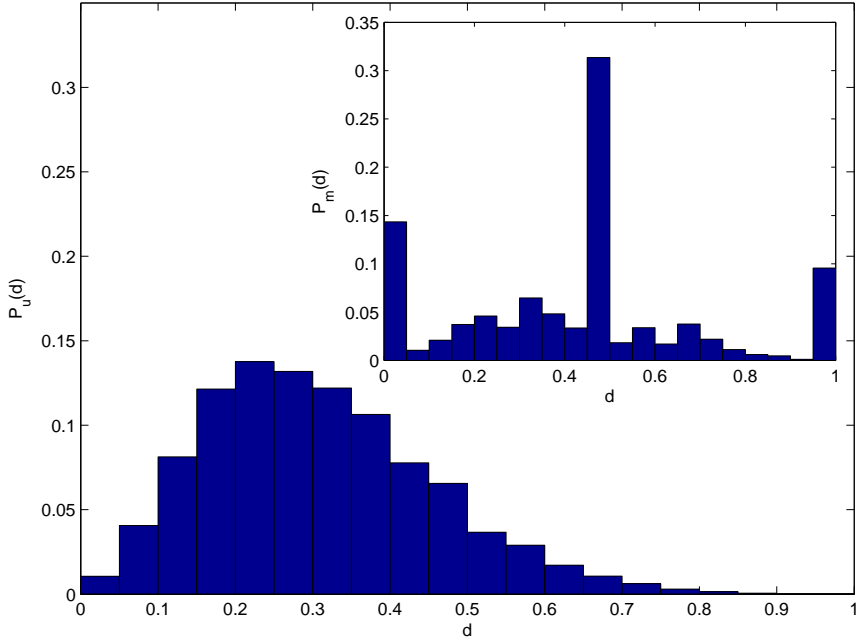}
\caption{\label{fig:three}Distribution $P_{u}(d)= \sum_{i}\sum_{j \ne i}\delta(d_{ij},d)/N(N-1)$ of
distance between users and the distribution
$P_{m}(d)= \sum_{\alpha}\sum_{\beta \ne \alpha} \delta(d_{\alpha\beta},d)/M(M-1)$ of distance between
movies. $\delta(d_{x},d)$ is the Kronecker symbol, $N$ is the
total number of people in the population and $M$ is the total number of movies.}
\end{figure}

Duality picture. The voting matrix $V$ can be viewed in two ways. In
user-centric view we measure the pairwise affinity of users.
The affinity distribution indicates how much information redundancy
is buried in the data to predict users' opinion about a movie. This
is similar to Newman's 'Ego-centered networks' \cite{Newman01}. In the
movie-centric view we look at the distribution of movie affinity.
This shows how controversial movies were voted by the
population. This ``duality picture'' is not
symmetric Fig.(\ref{fig:three}).

Let us start with the user-centric view. We define the overlap
between users $i$ and $j$ as
\begin{equation}
\Omega_{ij}= \frac{ \sum_{\alpha=1}^M v_{i\alpha}
v_{j\alpha}}{\sum_{\alpha=1}^M |v_{i\alpha}| |v_{j\alpha}|}, \quad
\Omega_{ij} \in (-1,1). \label{eq:one}
\end{equation}
This measures the affinity between users $i$ and
$j$. $\Omega_{ij}$ close to $1$ means similar tastes, whereas
$\Omega_{ij}$ close to $-1$ means opposite tastes.
$\sum_{\alpha=1}^M |v_{i\alpha}| |v_{j\alpha}|$ gives the number of
commonly seen movies by both users $i$ and $j$. $|\cdot|$ denotes
the absolute.

In the movie-centric view the affinity between two movies is defined
in an analogous way as follows:
\begin{equation}
\Omega_{\alpha\beta}= \frac{ \sum_{i=1}^N v_{i\alpha}
v_{i\beta}}{\sum_{i=1}^N |v_{i\alpha}| |v_{i\beta}|}, \quad
\Omega_{\alpha \beta} \in (-1,1). \label{eq:two}
\end{equation}
$\Omega_{\alpha\beta}$ close to $1$ means that movie $\alpha$ and
movie $\beta$ are judged as similar by each user, whereas
$\Omega_{\alpha\beta}$ close to $-1$ indicates that the two movies
are judged to be opposite. $\sum_{i=1}^N |v_{i\alpha}| |v_{i\beta}|$
gives the number of people who have seen both movies $\alpha$ and
$\beta$.
A more intuitive concept is given by the distance $d_{ij} = (1
-\Omega_{ij})/2$ for users and $d_{\alpha\beta} = (1
-\Omega_{\alpha\beta})/2$ for movies respectively. $d_{ij} \sim 0$
represents similar tastes for user $i$ and user $j$ whereas $d_{ij}
\sim 1$ opposite opinions. Likewise interpretations for the
movie-centric view.

$P_u(d)$ in Fig.(\ref{fig:three})indicates a rather homogenous
distribution of tastes among users. Furthermore the peak around $d
\sim 0.2$ implies a rich information source which allows taste
prediction. If users would vote in a random manner the peak would be
around $0.5$. On the other hand in the movie-centric view the
distribution $P_{m}(d)$ in Fig.(\ref{fig:three}) appears more
polarized. One explanation for this is the following: the overlaps of
the users are typically averaged over a lot of bits (from every user there
are at least 200 opinions known), while many movies are only
few times voted. Hence it is much easier to get a ``perfect'' $+1$ or
$-1$ overlap. Apart from this we observed two effects which also give hints about the
asymmetry between the two views. One example: for a \emph{Star wars}
movie the set of 'antipodes'- movie with $d \sim 1$ includes A) some
movies oriented for the audience of young women (e.g. \emph{Mr.
Wrong}); B) Less successful sequels of the \emph{Star Wars} trilogy
hated by some of their fans. It is not surprising that for movies of
type B there exists a considerable number of people who saw both of
them. What is more surprising is that for some of the movies of type
$A$ the number of users liking \emph{Star Wars} could also be quite
large. We tentatively attribute it to the `girl-friend effect' in
which \emph{Star Wars} fans were dragged by their girlfriends to see
a movie like \emph{Mr. Wrong}. Most of them disliked it (hence the
distance between these movies is close to $1$ in spite of a
relatively large common audience).

One can use the information of distances between movies to make
a proposition to users: if user $i$ likes movie $\alpha$
($v_{i\alpha}=1$) and this movie is within a distance $d\sim 0$ with 
movie $\beta$ it is very likely that user $i$ also will like movie
$\beta$.

However to predict a vote $v_{i\alpha}$ we
will use the information of affinity between users. Here, user $i$ is the
'center' of the universe and all others have certain distances to him. Users
close to him are more trustful because they share similar tastes. Hence
they should have more weight in the prediction.
Furthermore we have to penalize users who have not seen that much movies in
common. In this way we take care of the statistical significance.

We introduce our method to predict votes: the dataset of votes (matrix $V$) is
divided into a 'training' set $V_{train}$ and a 'test' set $V_{test}$. The votes of the two sets
are generated randomly out of the voting matrix $V$. The votes in $V_{train}$ are treated as
observed whereas the votes in $V_{test}$ are hidden for the algorithm. That is we use
votes in $V_{train}$ to predict votes in $V_{test}$.

For prediction we use the following form
\begin{equation}
v'_{i\alpha} = \frac{ \sum_{j \ne i} \Omega_{ij} \sqrt{\omega_{ij}}}
{ \sum_{j  \ne i} |\Omega_{ij}| \sqrt{\omega_{ij}}} \cdot \frac{v_{j_ \alpha}}{|v_{j \alpha}|}.
\label{eq:three}
\end{equation}
Where $v'_{i\alpha}$ is the predicted vote which has to
be compared to $v_{i\alpha} \in V_{test}$. $v_{j\alpha} \in V_{train}$ are the votes
which are supposed as known . Statistical significance is taken
into account by $\omega_{ij}= \sum_{\alpha} |v_{i\alpha}| |v_{j\alpha}|$,
which is the number of shared movies between user $i$ and user $j$.
Our measure of accuracy is given by
\begin{equation}
s_{i\alpha}=\left\{ \begin{array}{ll}
1 & \textrm{ if $sign(v'_{i\alpha}) = v_{i\alpha}$}\\
0 & \textrm{ otherwise}
\end{array} \right.,  \Pi=\frac{\left( \sum_{i}\sum_{\alpha} s_{i\alpha}
    \right)}{|V_{test}|}
\label{eq:four}
\end{equation}
Where $\Pi \in (0.5,1.0)$, $|V_{test}|$ is the number of
votes we want to predict and $v_{i\alpha} \in V_{test}$.

$\Pi \sim 0.5$ means no predictive power. In this case prediction is
random whereas $\Pi=1$ gives an accuracy of $100\%$ (every vote was predicted
correctly).

\begin{figure}
\includegraphics[width=0.9 \textwidth]{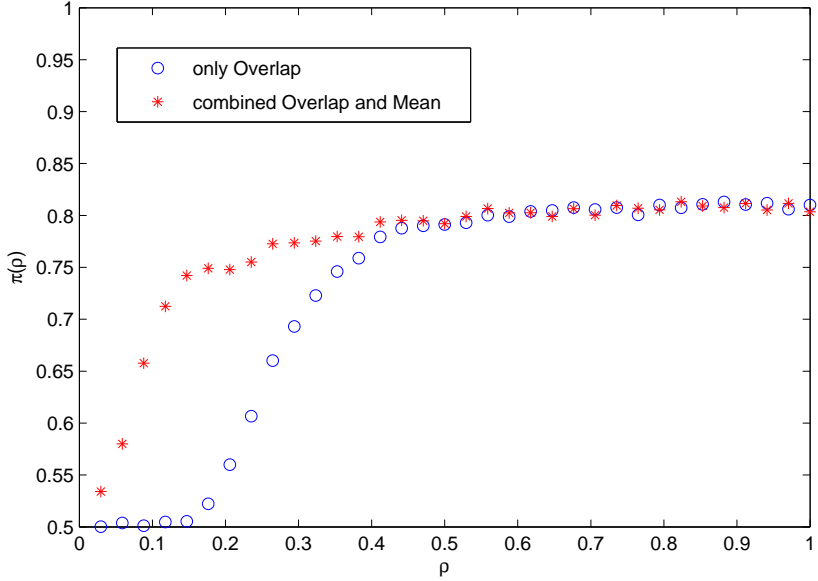}
\caption{\label{fig:one}The prediction power $\Pi(\rho)$ as a function
of $\rho$. $\rho$ is  the fraction of present votes in $V_{train}$
to the total number of votes in the voting matrix $V$.}
\end{figure}

It is a common belief that prediction accuracy in 'recommender
systems' is an increasing function of the available amount of data. The more votes the better. 
However, our result shows a saturation
of the prediction power  after a critical mass of
data Fig.(\ref{fig:one}).
We can clearly distinguish two phases. In the region $\rho \le 0.2$ no reasonable
prediction can be done, because there are not enough overlaps present. In this region the prediction
is by chance. By increasing the number of votes in $V_{train}$ - the prediction
accuracy increases too. However, after a critical value of $\rho \sim 0.6$ the
predictability saturates, without any further improvement with additional data input.
When we use somewhat different method with the mean tendance as an aide, that is
\begin{equation}
v'_{i\alpha} = \bar{v_{\alpha}} +\frac{ \sum_{j \ne i} \Omega_{ij} \sqrt{\omega_{ij}}}
{ \sum_{j  \ne i} |\Omega_{ij}| \sqrt{\omega_{ij}}} \cdot \frac{v_{j_ \alpha}}{|v_{j \alpha}|}.
\label{eq:five}
\end{equation}
the onset of the plateau is much earlier, in a sense this represents a big improvement. 
$\bar{v_{\alpha}} = \sum_{i} v_{i\alpha}/N_{\alpha}$
denotes the average vote of a movie $\alpha$ and $N_{\alpha}$ is the number of people
who voted for movie $\alpha$. However the plateau value remains the
same. This hints some fundamental limit at work, for this we need examine the origins of noise
intrinsically buried in the data. First of all, the massive collection of thousands web surfers is
far from being a precise process, an average user often votes carelessly, and with biases and whim,
typical of any human experiment. However if a rater sometimes votes random, and random data won't show
any meaningful correlation, as pointed out by \cite{Maslov01}, on the aggregate one must expect that there
is some coherence left in the data, its less-than-perfect collection quality finally shows up in our
calculation. It is remarkable that this degree of imperfection can be calculated at all. Though we should
never expect perfection in human endeavors, but significant room left for
improvement. Prediction quality can never attain $1$, no matter how good is the method and data \cite{Will94}.

\begin{figure}
\includegraphics[width=0.9 \textwidth]{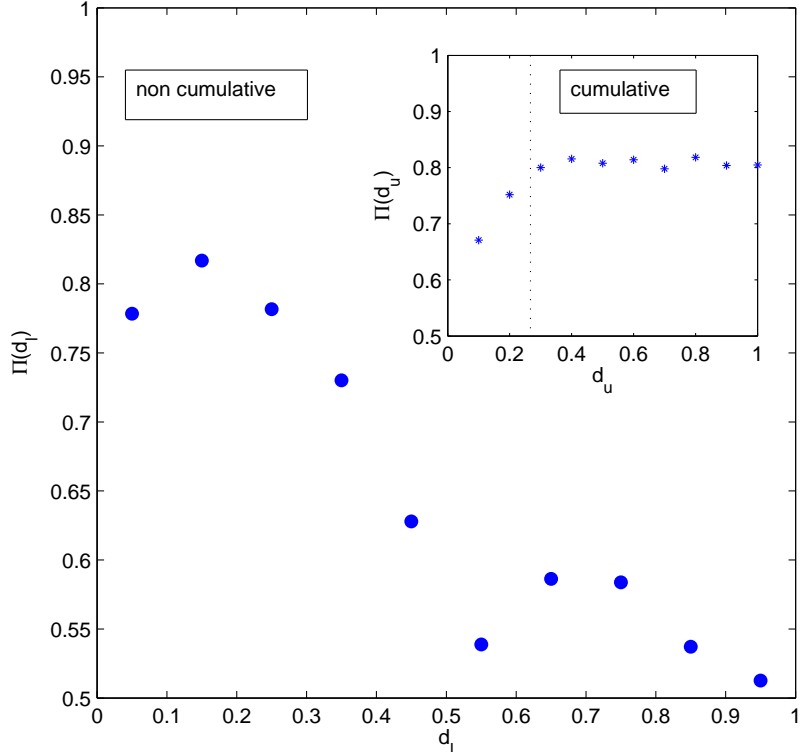}
\caption{\label{fig:two}The prediction power $\Pi(d_{l})$ as a function
of the lower distance threshold for the non cumulative case and $\Pi(d_{u})$ as a function of
the upper threshold (small box) for the cumulative case. Note that the
calculated accuracy for the non cumulative case is plotted always between the lower and the upper
distance threshold.}
\end{figure}

We investigate in more detail what are  crucial parameters for  prediction
accuracy. Fig.(\ref{fig:two}) shows a non cumulative and a cumulative plot of the prediction power.
In the non cumulative case we only take into account users within a certain
range of distance. Predicting $v_{i\alpha}$ (the vote from user
$i$ to movie $\alpha$) we build a subset of users $A^{(i)}_{d_{l}} = \{j \ne i| d_{l} \le d_{ij} \le d_{l} + 0.1 \}$ and
use only members of this set to predict votes in question. $d_{l} \in \{0.0,0.1 \cdots , 0.9\}$ is the
lower distance threshold. The upper distance threshold is given by $d_{l} + 0.1$.
Prediction power is given again by Eq.(\ref{eq:four}). 
For the cumulative case $d_{l}$ remains always $0$
and we vary only the upper distance threshold. We
build a subset of users $B^{(i)}_{d_{u}} = \{j \ne i| 0.0 \le d_{ij} \le d_{u} \}$ 
to predict vote $v_{i\alpha}$.
$d_{u} \in \{0.1,0.2, \cdots , 1.0\}$  denotes the upper distance threshold.
We observe in the non cumulative case that my `antipodes' Fig.(\ref{fig:two}) still could be
used for prediction (albeit poorly). However users who are very similar to 'me' are best
in predicting my tastes. The number of users within a small distance $d$ to a given user is low
but their predictions are good, while the number of users at intermediate distances $d\sim 0.3$ is
large but their predictive power is poor. One needs to strike a balance. As one can see in the
cumulative case Fig.(\ref{fig:two}) prediction power saturates around $d=0.2$ (indicated by the dotted line).
So there is no harm in including the votes from all users (provided that we weight them as we do).

\begin{figure}
\includegraphics[width=0.9 \textwidth]{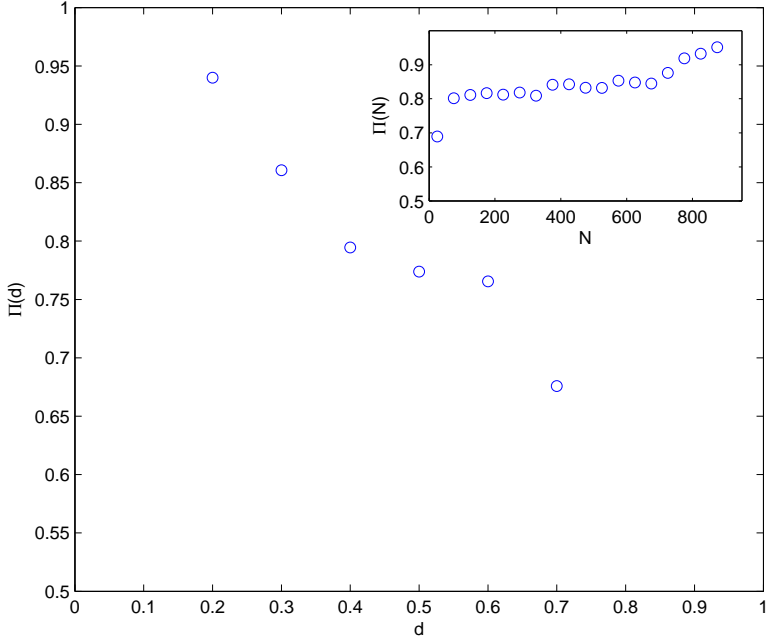}
\caption{\label{fig:four}The prediction power $\Pi(d)$ as a function
of the mean distance $d$  and the
predictability $\Pi(N)$ for movies as a function of the number of votes $N$ it has (small box).
The two plots are non cumulative. An example: $\Pi(0.2)$ gives the average predictability
of users who have an average distance $\bar{d_{i}}=\sum_{j \ne i} d_{ij}/(N-1) \le d=0.2$ to 
the rest of the population, $\Pi(0.3)$ gives the average predictability of users who have an average
distance $d= 0.2 \le \bar{d_{i}}=\sum_{j \ne i} d_{ij}/(N-1) \le d=0.3$ and so on. 
The plot for the movie predictability (small box) is also non cumulative and indicates 
an increasing prediction accuracy for an increasing number of votes.}
\end{figure}

Next we investigate what determines the mean predictability of a user or a movie Fig.(\ref{fig:four}).
People who have a small average distance $\bar{d_{i}}=\sum_{j \ne i} d_{ij}/(N-1)$ to the rest of the
population are better predictable then people who have somewhat special tastes.
If somebody follows the mainstream he or she will have more users with similar tastes which
are best for predictions. Note that the predictability seems to extrapolate
to $1$ for small $d$.

The major determinant of predictability of a movie is how many votes it has. This is quantified
in Fig.(\ref{fig:four}). It could be interpreted like this: the prediction of an opinion of a given
user on a popular movie could be based on large ensemble of other users who also saw this movie. Chances
are that this ensemble would contain decent number of users with tastes similar to the user we are
currently trying to predict. Thus the prediction would turn out to be more precise.
\section{Conclusion}
To conclude we note that our relatively straightforward method can
yield significant prediction precision. However there seems to have
an intrinsic limit in the precision that should be attributed to the
original noisy source. Our results reveal that people's tastes tend
to be homogenous whereas movies are polarized. The implications of
our study go much beyond merely predicting user's tastes. One can
image that consumers' relation with myriad of products and services
as a much larger information matrix. It would have significant
impact on the economy if a consumer's potential tastes to the vast
majority of products and services that she has not yet tested can,
to a reasonable precision, be predicted. With the rapid evolution of
the Information Technology, where the feedbacks from consumers can
be effectively tracked and analyzed, it is not to far-fetched to see
our economy completed transformed by a new paradigm.
\bibliographystyle{elsart-num}
\bibliography{article_movielens}
\end{document}